\documentclass[aps,preprint,onecolumn,secnumarabic,nobalancelastpage,amsmath,amssymb,
nofootinbib]{revtex4}

\usepackage{pst-plot}
\usepackage{pstricks,pstricks-add}
\usepackage[scriptsize,nooneline,hang]{caption}      
\usepackage[hang,nooneline,scriptsize]{subfigure}
     
\usepackage{graphics}
\usepackage{graphicx}     
\usepackage{url}
\usepackage{dcolumn}         
\usepackage{hyperref} 
\usepackage{bm}               
\usepackage{mathrsfs}
\usepackage{epstopdf}
\usepackage{hyperref}         
\usepackage{color}

\begin{document}

reprint{APS/123-QED}

\title{Shadow of a Kerr-Sen dilaton-axion Black Hole}

\author{Sara Dastan}
\author{Reza Saffari}
\email{rsk@guilan.ac.ir}
\author{Saheb Soroushfar}
 \affiliation{Department of Physics, University of Guilan, 41335-1914, Rasht, Iran.}%



\date{\today}

\begin{abstract}
We analiyze the shadow of charged stationary axially symmetric space-time (Kerr-Sen dilaton-axion black hole).
This black hole is defined by a mass $M$, a spin $a$ and $r_{\alpha}=\frac{Q^{2}}{M}$, where $Q$ is the electric charge. Shadows are investigated in two conditions, i) for an observer at infinity in vacuum and ii) for an observer at infinity in the presence of plasma with radial power-low density. In vacuum, the shadow of this black hole depends on charge and spin parameter. We can see that, increasing electric charge $Q$, decreases the size of shadow. Also, increasing spin parameter $a$, decreases the size of shadow. However, in existence of plasma, parameter of plasma like refraction index, playing an  important role on shadows. In fact, decreasing refraction index $n$, decreases the size of shadow.
\end{abstract}

\maketitle

\section{INTRODUCTION}
A black hole is a part of space-time, which displays powerful gravitational effects that nothing, including particles and electromagnetic radiation, such as light, can escape from inside it~\cite{Wald:1984}.
Also, detecting black holes is an interesting issue, which attracts lots of researchers and scientists, who provide lots of articles and literatures. However, lack of observational data is one of the most important problems in gathering information about black holes.

In this context, General Relativity predicts two methods for observing black hole, which are called gravitational lensing~\cite{Gyulchev:2009dx}
,~\cite{Gyulchev:2006zg} and detection of gravitational waves~\cite{Antonucci:2011zza}.
Nowadays, obtaining data about \textit{Sagittarius A*}, a possible black hole in the center of the Milky Way, is an interesting subject. As a result, two groups were formed with the objective to see the possible black hole in the center of the Milky Way, these groups are known as \textit{Event Horizon Telescope} (EHT) ~\cite{Doeleman:2008qh,EHT:2008qh} and \textit{BlackHoleCam} (BHC) ~\cite{BHC:2008qh}. The goal of these groups is detecting the shadow of the possible black hole in the center of the Milky Way.
The concept of shadow is explained in following. If a source of light and an observer placed on $r_{L}$ and $r_{O}$ from a black hole respectively, with the term of $r_{L}>r_{O}$, the sky of observer has two condition. First, the sky of observer will be dark if the light rays go to the event horizon of black hole and they do not come back to the light source. On the other hand, the sky of observer will be bright, when the light rays are deflected by the black hole and they come back to the light source. \textbf{Shadow} is the darkness of the observer's sky~\cite{Grenzebach:2014fha}.
The shadow of different space-time has been analyzed in various articles. In fact, investigating the black hole's shadow, can improve our information about it. Surveys show that, the shadow of a Schwarzschild black hole is circular and it has photon sphere~\cite{synge}, while the shadow of a Kerr black hole is not circular and it has photon region~\cite{Chandrasekhar:1985kt}. Some of the study of the black hole's shadow are, Kerr~\cite{Bardeen,Cunha:2015yba,Vincent:2016sjq}, Kerr- Newman~\cite{kerr-newman}, Randall-Sundrum braneworld~\cite{Amarilla:2011fx}, black holes in $f(R)$ gravity~\cite{Dastan:2016vhb}, multi-black hole~\cite{Yumoto:2012kz} and black hole in extended Chern-Simons modified gravity~\cite{Amarilla:2010zq}.

In this paper, we come with the idea that, the possible black hole in the center of the Milky Way (\textit{Sagittarius A*}), is Kerr-Sen dilaton-axion black hole, which was introduced by Sen~\cite{Sen:1992ua} in 1992. The solution made of the classical equation of motion for heterotic string theory in the low energy effective field. It had been started from a Kerr black hole with twisting method, as a result, sometimes it is named twisted Kerr black hole. Some properties like, Thermodynamic~\cite{Larranaga:2010mb,Pradhan:2015yea}, astrophysical consequences~\cite{Jiang:2006mu,Li:2007af,Yang:2007ny,Chen:2009zzk}, hidden symmetries, photon capture~\cite{Hioki:2008zw} and geodesic motion~\cite{Soroushfar:2016yea} in Kerr-Sen dilaton-axion black hole have been studied. In this paper, we investigate the shadow of Kerr-Sen dilaton-axion black hole, for this purpose our paper is organized as follow, in Sect.~\ref{field}, we summarize the properties of Kerr-Sen dilaton-axion black hole, in Sect.~\ref{geodesic}, we study the geodesic equation of this black hole. We study the shadow for an observer at $r=\infty$ in Sect.\ref{infinite}. The shadow of black hole in the presence of plasma, has been analyzed in Sect.~\ref{plasma} and our results conclude in Sect.~\ref{conclusion}.

\section{Metric}\label{field}
 In this section, we study the charged, stationary, axially-symmetric solution \cite{Yazadjiev:1999ce} of
the field equations which was found by Sen. This solution was constructed using space duality and the classical Kerr solution. So, the line element in Boyer--Lindquist coordinates is written as,
\begin{align}
ds^{2}=-\big(1-\dfrac{2Mr}{\rho^{2}}\big)dt^{2}
+\dfrac{\rho^{2}}{\Delta}dr^{2}+\rho^{2}d\theta^{2} - \dfrac{4Mra\sin^{2}\theta}{\rho^{2}}dtd\varphi \nonumber\\
+\big(r(r+r_{\alpha})+a^{2}+\dfrac{2Mra^{2}\sin ^{2}\theta}{\rho^{2}}\big)\sin ^{2}\theta d\varphi^{2},
\end{align}
where
\begin{align}
\Delta_{r}=r(r+r_{\alpha})-2Mr+a^{2},
\end{align}
\begin{align}
\rho^{2}=r(r+r_{\alpha})+a^{2}\cos ^{2}\theta, \qquad
\end{align}
$M$ is the mass, $r_{\alpha} = \dfrac{Q^{2}}{M}$, where Q is the charge and $a= J/M$ is the angular momentum of the black hole. 
The Kerr-Sen black hole converts to the Gibbons-Maeda-Garfinkle-Horowitz-Strominger (GMGHS) black hole, when $a=0$, 
and in the case of $r_{\alpha}=0$ this black hole is Kerr. If both $a$ and $r_{\alpha}$ are equal to zero  then it is Schwarzschild black hole.

\section{The geodesic equations}\label{geodesic}
In this section, the geodesic equation of the Kerr-Sen dilaton-axion black holes are studied. The Hamilton--Jacobi equation
\begin{equation}\label{Hamilton}
\dfrac{\partial S}{\partial\tau} +\frac{1}{2} \ g^{ij}\dfrac{\partial S}{\partial x^{i}}\dfrac{\partial S}{\partial x^{j}}=0 
\end{equation}
can be solved with an ansatz for the action
\begin{equation}
\label{S}
S=\frac{1}{2}\varepsilon \tau - Et+L_{z}\phi +S_{\theta}(\theta) + S_{r} (r).
\end{equation}
The energy $E$ and the angular momentum $L$ which are given
by $P_{t}$ and $P_{\phi}$ are the constants of motion,
\begin{equation}\label{constants of motion}
P_{t}=g_{tt}\dot{t}+g_{t \varphi}\dot{\varphi}=-E,  \qquad P_{\phi}=g_{\varphi \varphi}\dot\varphi +g_{t \varphi}\dot{t} =L.
\end{equation}
Using Eqs.~(\ref{Hamilton})--(\ref{constants of motion}) we have
\begin{align}\label{ds/dr.ds/dtheta}
-\Delta_{r}\big(\dfrac{ds}{dr}\big)^{2} - \varepsilon r^{2}+\dfrac{1}{\Delta_{r}}\big((a^{2}+r(r+r_{\alpha}))E - aL\big)^{2} - \big(aE - L\big)^{2} = \nonumber\\
\big(\dfrac{ds}{d\theta}\big)^{2}+\varepsilon a^{2}cos^{2}\theta + \big(\dfrac{L^{2}}{sin^{2}\theta} - a^{2}E^{2}\big)cos^{2}\theta,
\end{align}
where, each side of Eq.(\ref{ds/dr.ds/dtheta}) depends only on $r$ or $\theta$. 
Using the Carter \cite{K} constant and the separation ansatz Eq.(\ref{S}) 
we have the equations of motion:
\begin{align}\label{Joda}
\rho^{4}(\dfrac{dr}{d\tau})^{2}=-\Delta_{r}(K+\varepsilon r(r+r_{\alpha}))+\big((a^{2}+r(r+r_{\alpha}))E-aL \big)^{2}=R(r),
\end{align}
\begin{align}\label{thetatho}
\rho^{4}(\dfrac{d\theta}{d\tau})^{2}=\Delta_{\theta}(K-\varepsilon a^{2}cos^{2}\theta)-\dfrac{1}{sin^{2}\theta}\big(aE sin^{2}\theta -L \big)^{2} =\Theta(\theta),
\end{align}
\begin{align}
\rho^{2}(\dfrac{d\varphi}{d\tau})=\dfrac{a}{\Delta_{r}}\big((a^{2}+r(r+r_{\alpha})) E - aL \big)-\dfrac{1}{sin^{2}\theta}(a E sin^{2}\theta - L),
\end{align}
\begin{align}\label{ttho}
\rho^{2}(\dfrac{dt}{d\tau})=\dfrac{(a^{2}+r(r+r_{\alpha}))}{\Delta_{r}}\big((a^{2}+r(r+r_{\alpha})) E - aL \big) - a (a E sin^{2}\theta - L).
\end{align}
For studying the shadow null geodesic is important, so we put $\varepsilon$=0 and we have
\begin{align}\label{Joda.1}
\rho^{4}(\dfrac{dr}{d\tau})^{2}=-\Delta_{r} K+\big((a^{2}+r(r+r_{\alpha}))E-aL \big)^{2}=R(r),
\end{align}
\begin{align}\label{thetatho.1}
\rho^{4}(\dfrac{d\theta}{d\tau})^{2}=\Delta_{\theta} K-\dfrac{1}{sin^{2}\theta}\big(aE sin^{2}\theta -L \big)^{2} =\Theta(\theta),
\end{align}
\begin{align}
\rho^{2}(\dfrac{d\varphi}{d\tau})=\dfrac{a}{\Delta_{r}}\big((a^{2}+r(r+r_{\alpha})) E - aL \big)-\dfrac{1}{sin^{2}\theta}(a E sin^{2}\theta - L),
\end{align}
\begin{align}\label{ttho.1}
\rho^{2}(\dfrac{dt}{d\tau})=\dfrac{(a^{2}+r(r+r_{\alpha}))}{\Delta_{r}}\big((a^{2}+r(r+r_{\alpha})) E - aL \big) - a (a E sin^{2}\theta - L).
\end{align}
Introducing dimensionless quantities i.e. $\xi=\frac{L}{E}$ and $\eta=\frac{K}{E^{2}}$, which are constant along the geodesics, Eq.~(\ref{Joda.1}) becomes as,
\begin{align}
\rho^{4}(\dfrac{dr}{d\tau})^{2}=-\Delta_{r} (\eta E^{2})+\big((a^{2}+r(r+r_{\alpha}))E-a(\xi E) \big)^{2}=R(r)
\end{align}
For circular motion of the particles, the effective potential is useful tool that needs to be investigated, so we obtain~\cite{Wei:2013kza}
\begin{align}\label{18.3}
\rho^{4}(\dfrac{dr}{d\tau})^{2}+V_{eff}=0.
\end{align}
As a result,
\begin{align}\label{18.4}
V_{eff}=\Delta_{r} (\eta E^{2})-\big((a^{2}+r(r+r_{\alpha}))E-a(\xi E) \big)^{2}.
\end{align}
For finding $\xi$ and $\eta$, we analyze circular orbits of the photons \cite{Bardeen:1972fi}
\begin{equation}\label{18.5}
V_{eff}=0,  \qquad \frac{dV_{eff}}{dr}=0.
\end{equation}
These conditions are equivalent to $R(r)=0$ and $\dfrac{dR(r)}{dr}=0$. So, we can obtain the constant of motion i.e. $\eta$ and $\xi$, using Eqs.~(\ref{18.4}) and (\ref{18.5}).

\section{Shadow for an observer at $r\rightarrow\infty$}\label{infinite}
In this section, we analyze the shadow of black holes for an observer, which is placed in $r=\infty$ in vacuum, so the celestial coordinate $\alpha$ and $\beta$ are introduced as~\cite{Vazquez:2003zm}
\begin{equation}\label{alpha}
\alpha =\lim_{r_{O}\longrightarrow\infty} (r^{2}_{O} \sin (\theta_{O})\dfrac{d\varphi}{dr}), 
\end{equation}
and
\begin{equation}\label{beta}
\beta =\lim_{r_{O}\longrightarrow\infty} r^{2}_{O}\dfrac{d\theta}{dr},
\end{equation}
where, $\theta_{O}$ is the angle between the rotation axis of the black hole and the line of sight of the observer.
Considering the situation $r_{O}\rightarrow \infty$ and $\theta_{O}=\dfrac{\pi}{2}$ and using Eq.~(\ref{Joda.1})--~(\ref{ttho.1}) , $\alpha$ and $\beta$ are equal to
\begin{equation}\label{22.1}
\alpha = -\xi ,
\end{equation}
and
\begin{equation}\label{23.1}
\beta = \pm \sqrt{\eta}.
\end{equation}
By above conditions, for obtaining the black hole's shadow, we plot $\beta$ versus $\alpha$. In Fig.~\ref{infinity} different values of the rotation parameter ($a$ ) and electric charge ($Q$) are represented. We show that, by changing $a$ and $Q$, the symmetry and size of the black hole's shadow, will change. In fact, the size of shadow decreases by increasing $Q$ and the shadow's symmetry of the black hole, decreases by increasing $a$.
\begin{figure}[h]
	\centering
         \subfigure[]{
         \includegraphics[width=0.4\textwidth]{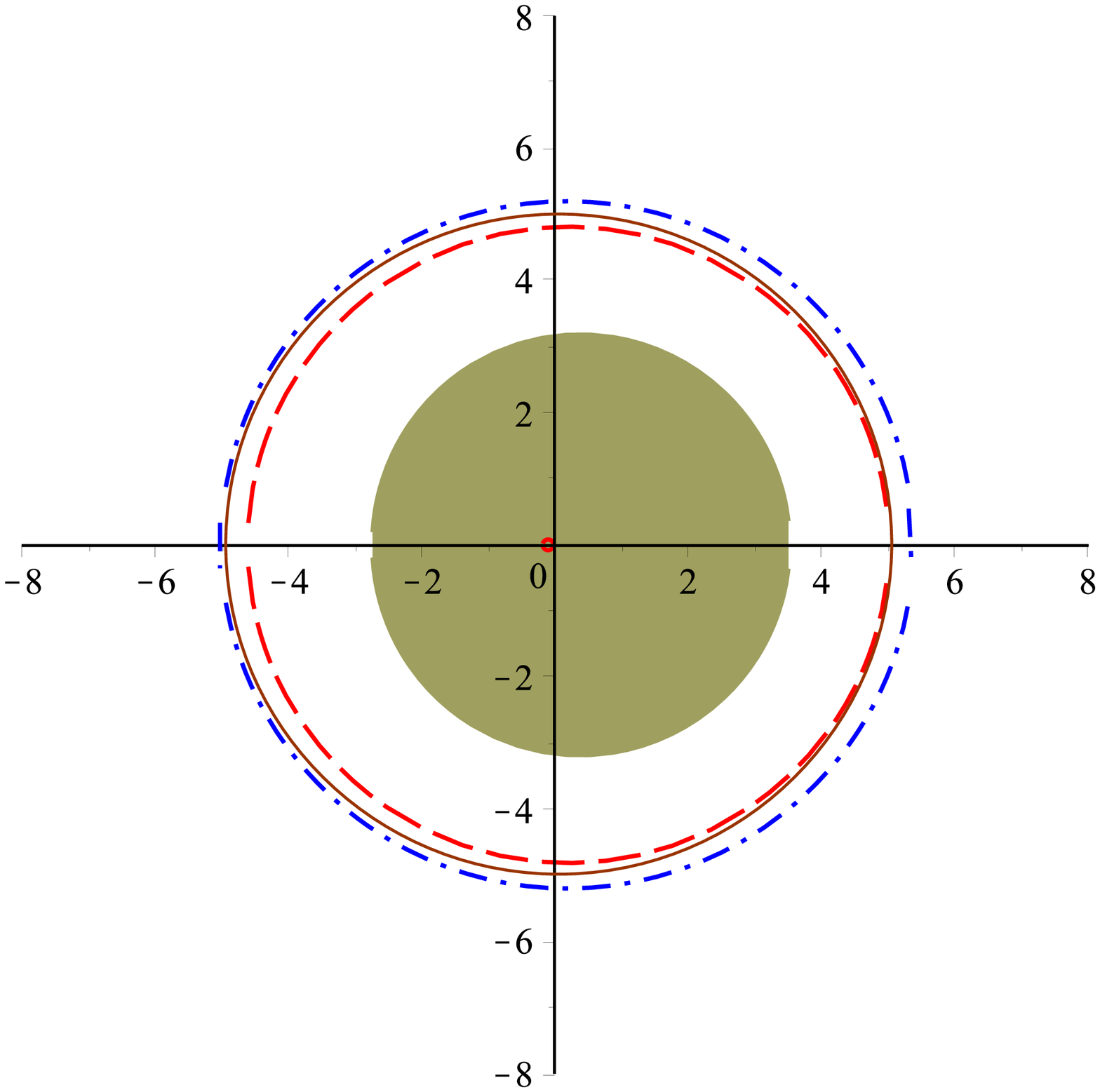}
	}
	\subfigure[]{
		\includegraphics[width=0.4\textwidth]{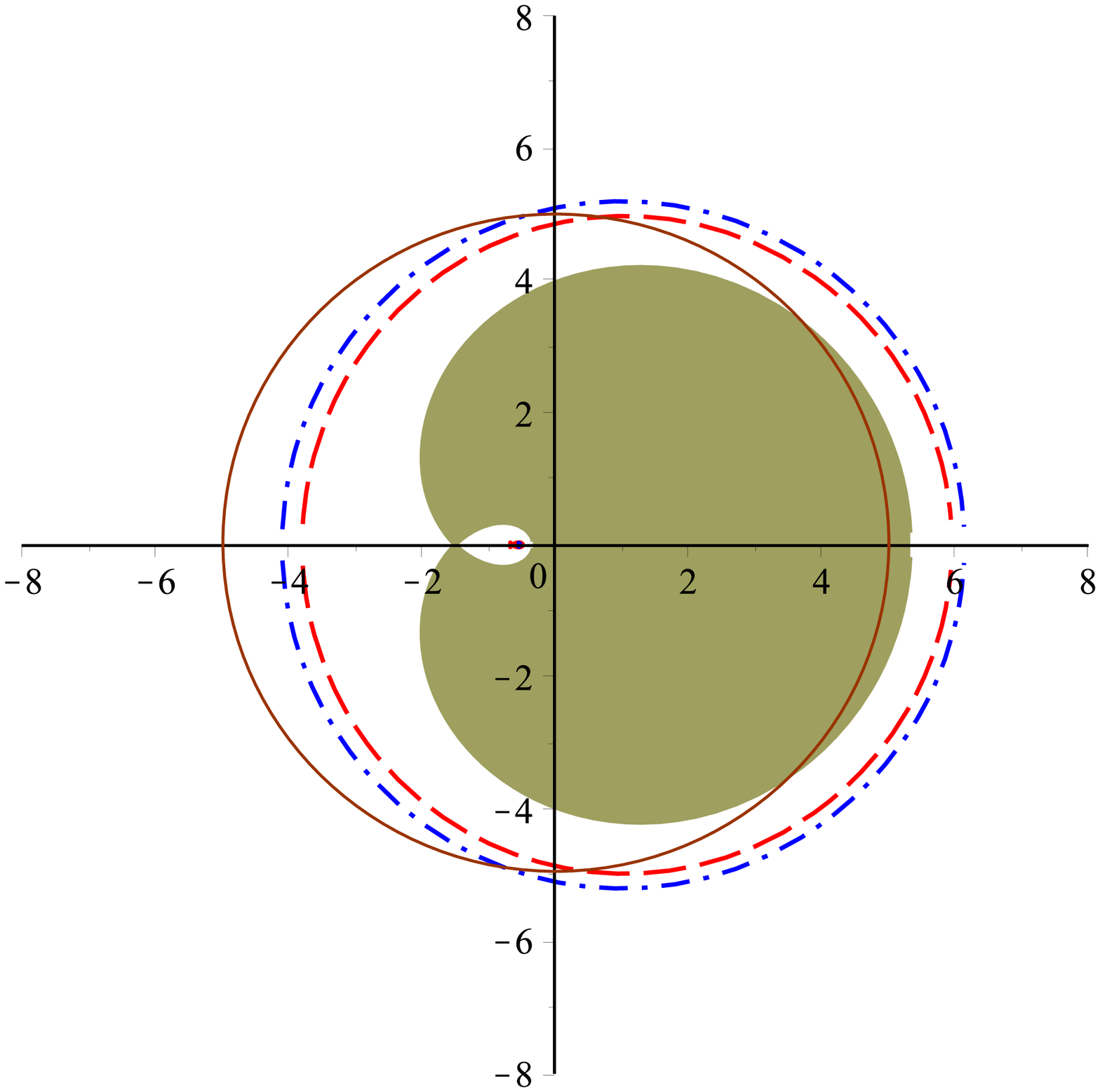}
	}
	\subfigure[]{
		\includegraphics[width=0.4\textwidth]{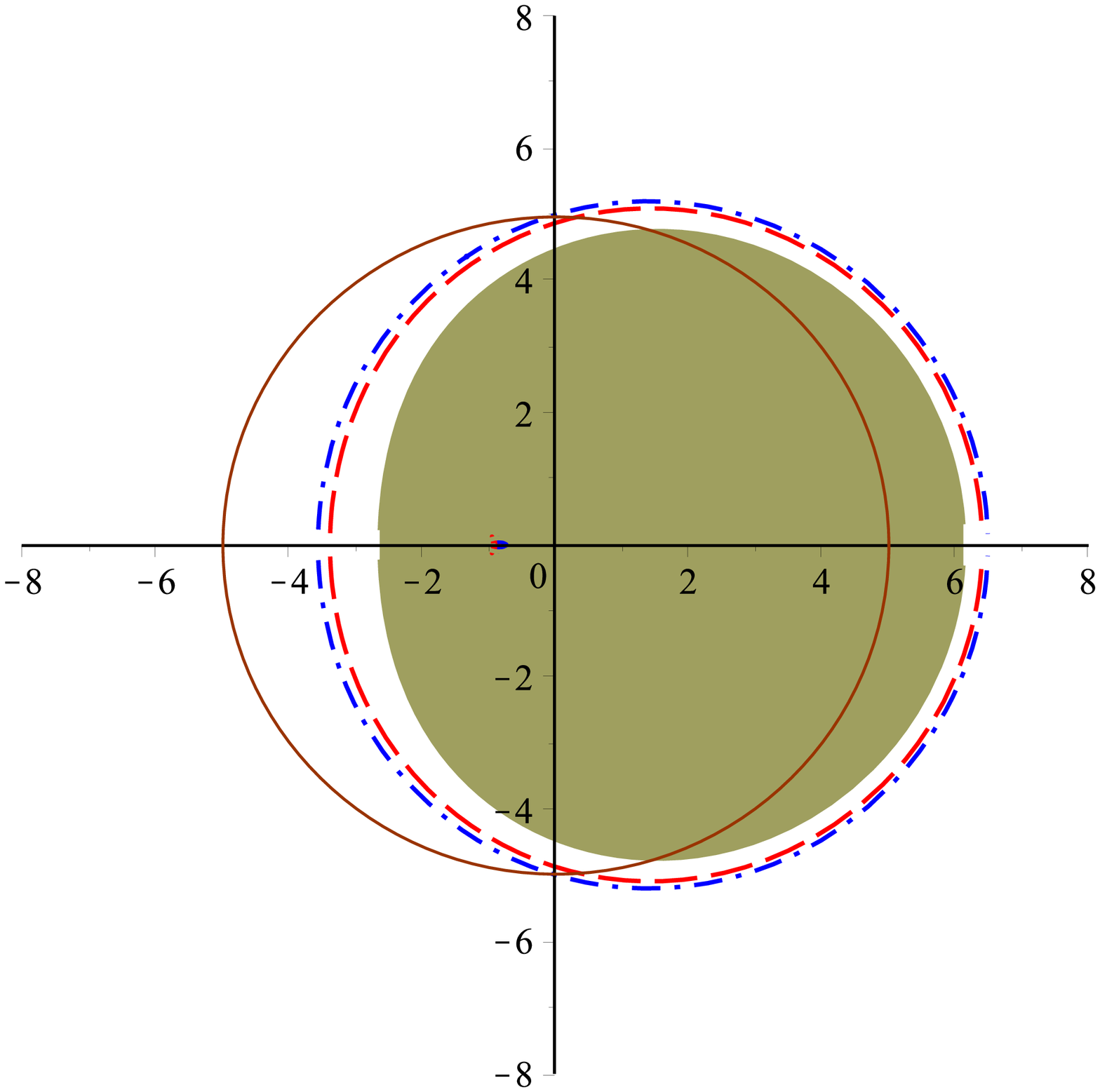}
	}
	\subfigure[]{
		\includegraphics[width=0.4\textwidth]{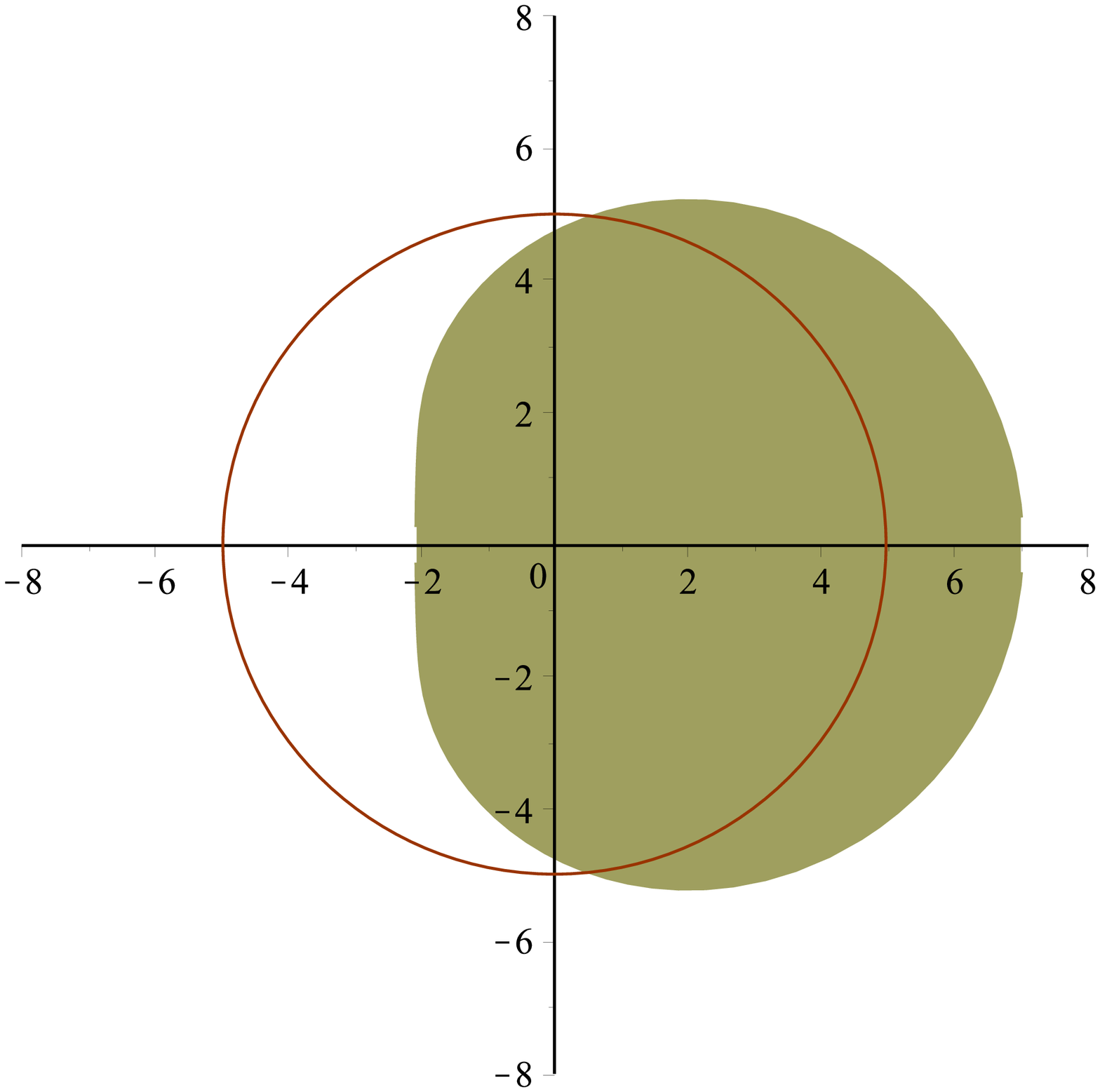}
	}
	\caption{\label{infinity}The image of black hole's shadow in Kerr-Sen Dilaton-Axion black hole for different values of rotation parameters, $a=0$, $a=0.5$, $a=0.7$ and $a=1$ for (a), (b), (c) and (d) respectively. For each value of $a$, $Q=0$(blue dashed-dotted line), $Q=\frac{Q_{c}}{2}$(red dashed line) and $Q=Q_{c}$(green filled solid line) for (a), (b), (c), but for (d) only, $Q=0$ is considered. The brown solid line is reference circle. The detail of parameters are shown in table~\ref{tab0} (see Appendix).}
	\label{infinity}
 \label{pic:shadow}
\end{figure}

In following, using the parameters which are introduced by Hioki and Maeda~\cite{Hioki:2009na}, we analyze deviation from circular form ($\delta_{s}$) and the size ($R_{s}$) of the shadow image of the black hole.
For calculating these parameters, we consider five points $(\alpha_{t},\beta_{t})$,$(\alpha_{b},\beta_{b})$, $(\alpha_{r},0)$, $(\alpha_{p}, 0)$ and  $(\bar{\alpha}_{p}, 0)$ which are top, bottom, rightmost, leftmost of the shadow and leftmost of the reference circle (see Fig~.\ref{deviation}~\cite{Abdujabbarov:2015xqa}) respectively, so we have 
  \begin{equation}\label{23.1}
 R_{s}=\dfrac{(\alpha_{t}-\alpha_{r})^{2}+\beta^{2}_{t}}{2(\alpha_{t}-\alpha_{r})}.
\end{equation}
and
\begin{equation}\label{23.2}
\delta_{s}=\dfrac{(\bar{\alpha}_{p}-\alpha_{p})}{R_{s}},
\end{equation}
In Fig.~\ref{RS}, we show the plot of $R_{s}$ and $\delta_{s}$ for different values of $Q$. We show that, by increasing $Q$, $\delta_{s}$ increases and $R_{s}$  decreases. 
It means that, the larger value of electric charge leads to decreasing in the size and increasing in the distortion of the shadow of black hole.
\begin{figure}[h]\label{deviation}
	\centering
	{
		\includegraphics[width=0.4\textwidth]{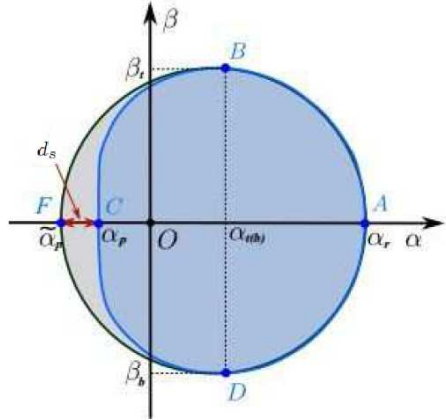}
	}
	\caption{\label{deviation}The black hole shadow and reference circle. $d_{s}$ is the distance between the left point of the shadow and the reference circle.}
 \label{pic:shadow}
\end{figure}

\begin{figure}[h]\label{RS}
	\centering
	\subfigure[$R_{s}$ and $Q$ for $\theta_{o}=\frac{\pi}{2}$ ]{
		\includegraphics[width=0.4\textwidth]{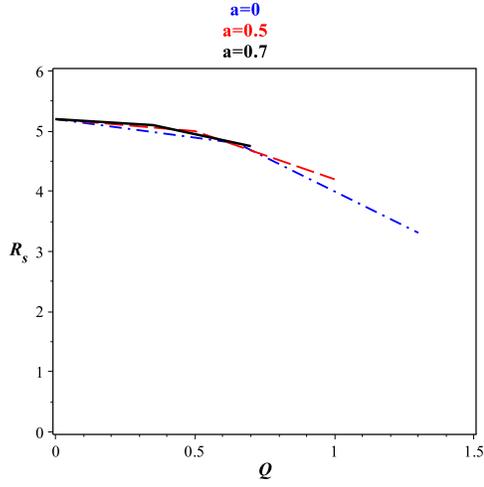}
	}
	\subfigure[$\delta_{s}$ and $Q$ for $\theta_{o}=\frac{\pi}{2}$]{
		\includegraphics[width=0.4\textwidth]{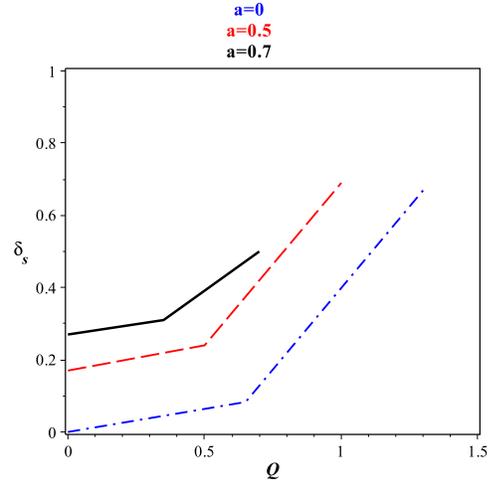}
	}
	\caption{\label{RS}Changes of $R_{s}$ and $\delta_{s}$ for an observer at $r \rightarrow \infty$. $a\simeq0$, $a=0.5$ and $a=0.7$ for the blue (dash-dotted line), the red (dashed line) and the black (solid line) respectively.}
 \label{pic:shadow}
\end{figure}
\clearpage
\subsection*{Energy emission rate}
In this part, we study the possible visibility of the Kerr-Sen dilaton-axion black hole through shadow. In the vicinity of limiting constant value, the cross section of the black hole's absorption moderate lightly at high energy. We know that a rotating black hole can absorb a electromagnetic waves, so the absorbing cross section for a spherically symmetric black hole is~\cite{Mashhoon:1973zz}
\begin{equation}\label{24}
\sigma_{lim}=\pi R^{2}_{s},
\end{equation}
using above equation~(\ref{24}), the energy emission rate is~\cite{Abdujabbarov:2016hnw}
\begin{equation}\label{25}
\frac{d^{2}E}{d\omega  dt}=\frac{2\pi R^{2}_{s}}{e^{(\frac{\omega}{T})}-1}\omega^{3},
\end{equation}
where $T$ is the Hawking temperature and $\omega$ the frequency. The Hawking temperature for this black hole, is obtained by the outer event horizon $r_{+}$ and mass of black hole $M$ as follow
\begin{equation}
r_{+}=M-\frac{Q^{2}}{2M}+\sqrt{(M-\frac{Q^{2}}{2M})^{2}-\frac{J^{2}}{M^{2}}},
\end{equation}

\begin{equation}\label{KMASSks}
M(S,Q,J)=\sqrt{\frac{4\pi^{2}J^{2}+2\pi Q^{2}S+S^{2}}{4\pi S}}.
\end{equation}

where $J=\frac{a}{M}$ and $S$ is the entropy of the black hole, which is equal to
\begin{equation}\label{ENks}
S=\pi (r^{2}_{+}+a^{2}).
\end{equation}
Using $ (T=\frac{\partial M}{\partial S}) $, the Hawking temperature is
\begin{equation}\label{KTks}
T=\frac{S^{2}-4\pi^{2}J^{2}}{4S\sqrt{4\pi^{3}J^{3}S+2\pi^{2}Q^{2}S^{2}+\pi S^{3}}}.
\end{equation}
Fig.~\ref{de} shows the plots of $\frac{d^{2}E}{d\omega dt}$ versus $\omega$. As one can see, energy emission rate decreases by increasing $Q$.
\begin{figure}[h]\label{de}
	\centering
	\subfigure[$a=0$]{
	\includegraphics[width=0.4\textwidth]{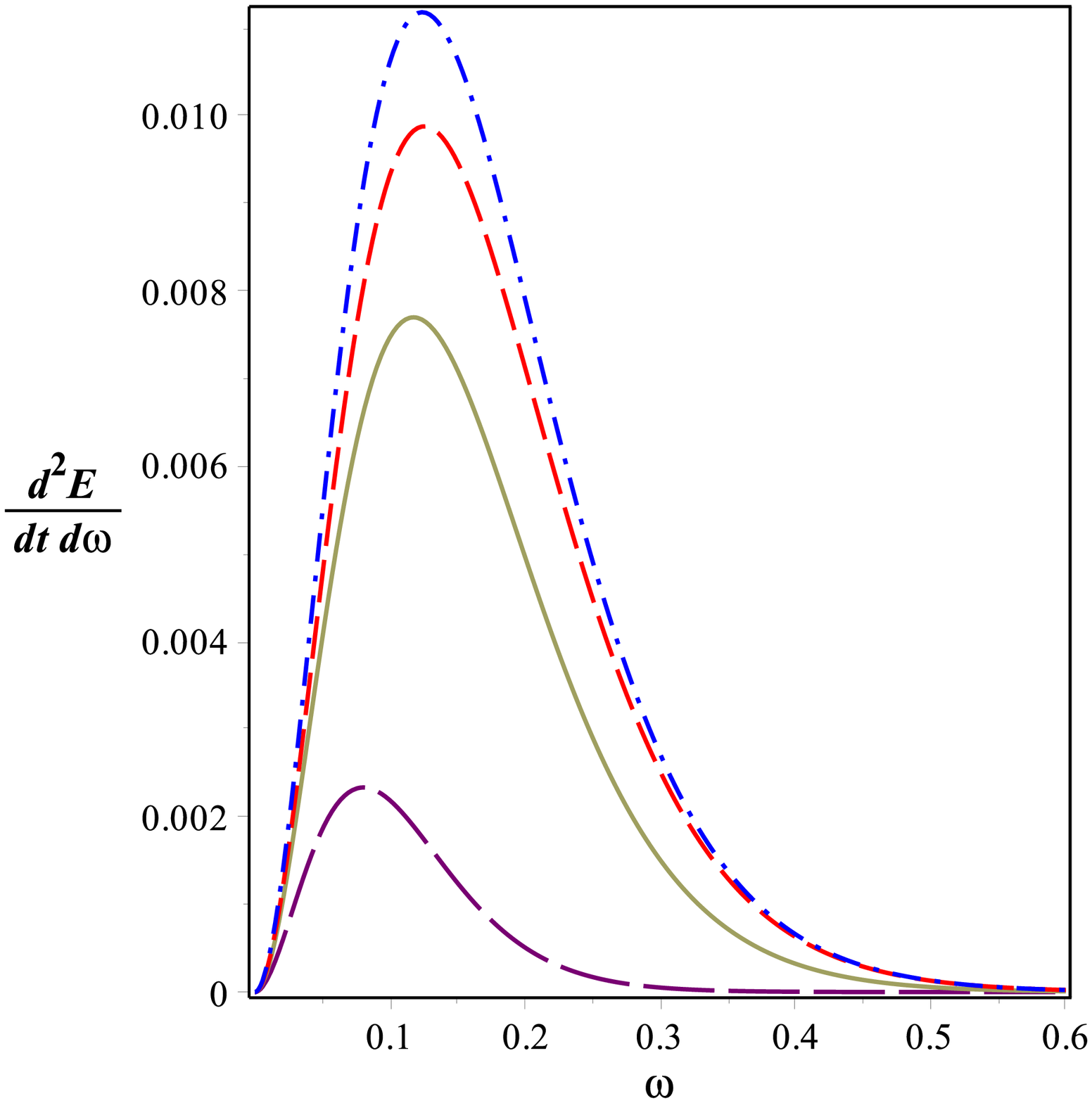}
	}
	\subfigure[$a=0.2$]{
		\includegraphics[width=0.4\textwidth]{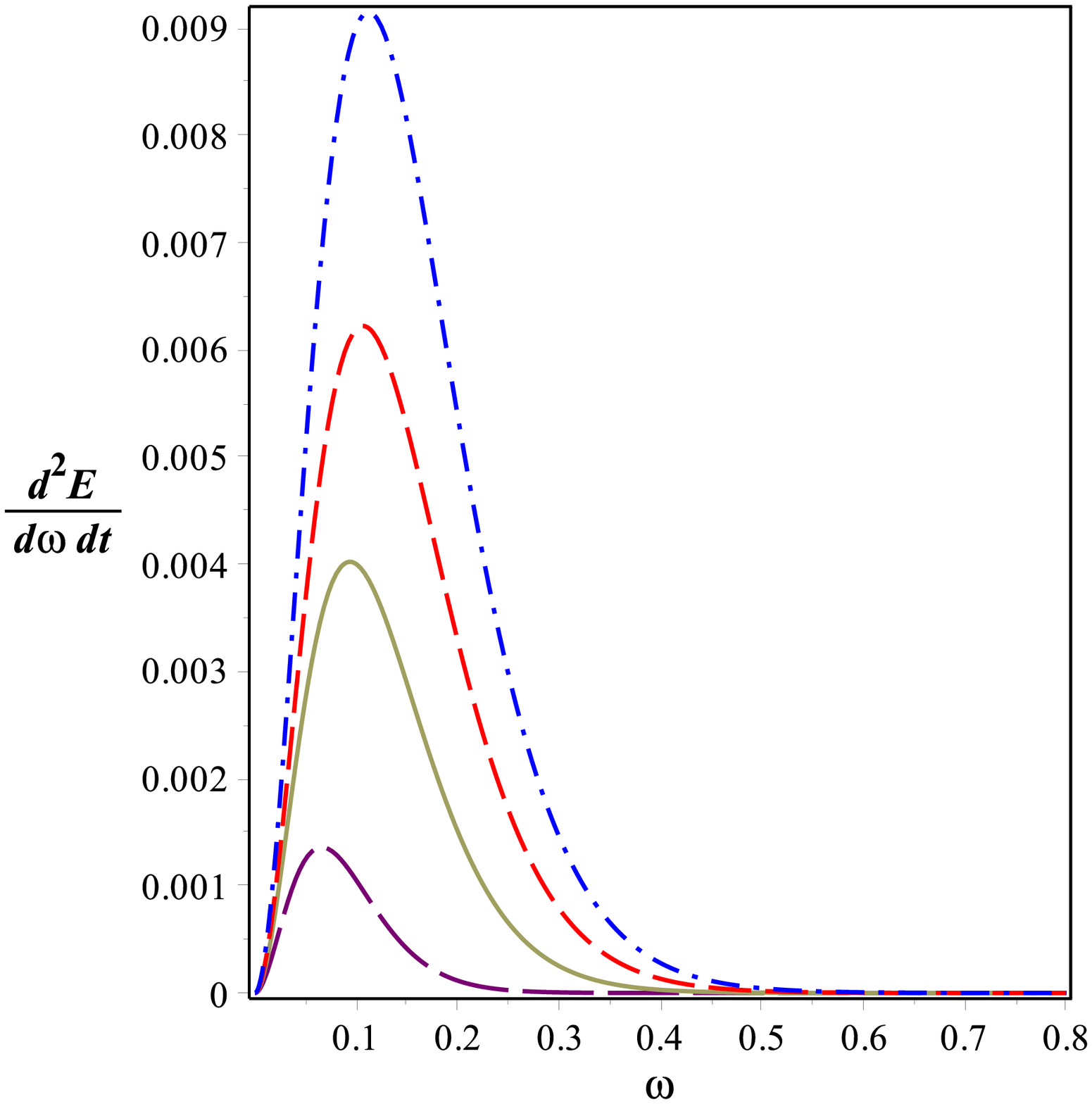}
}
	\subfigure[$a=0.5$]{
	\includegraphics[width=0.4\textwidth]{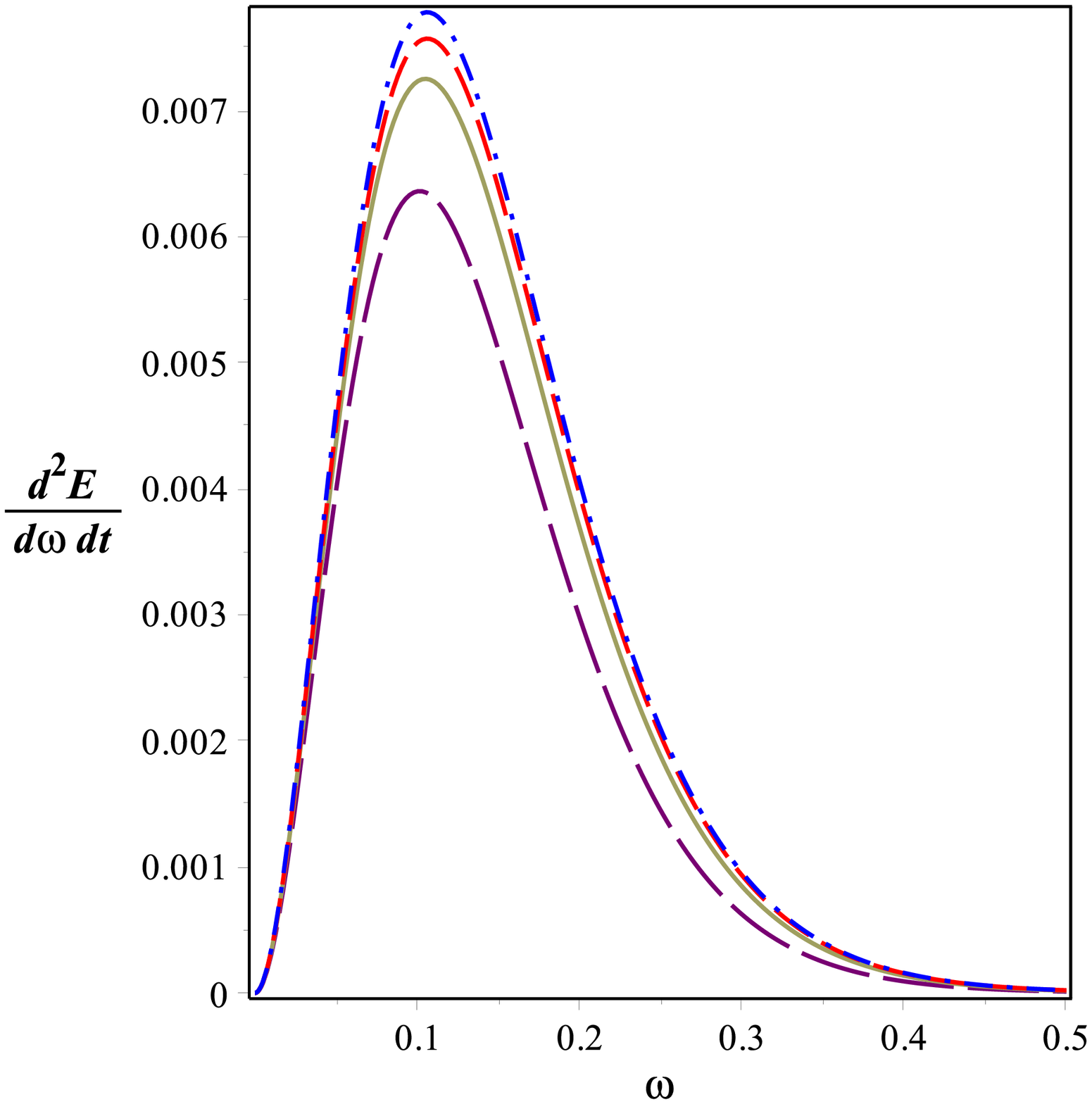}
	}
	\subfigure[$a=0.7$]{
	\includegraphics[width=0.4\textwidth]{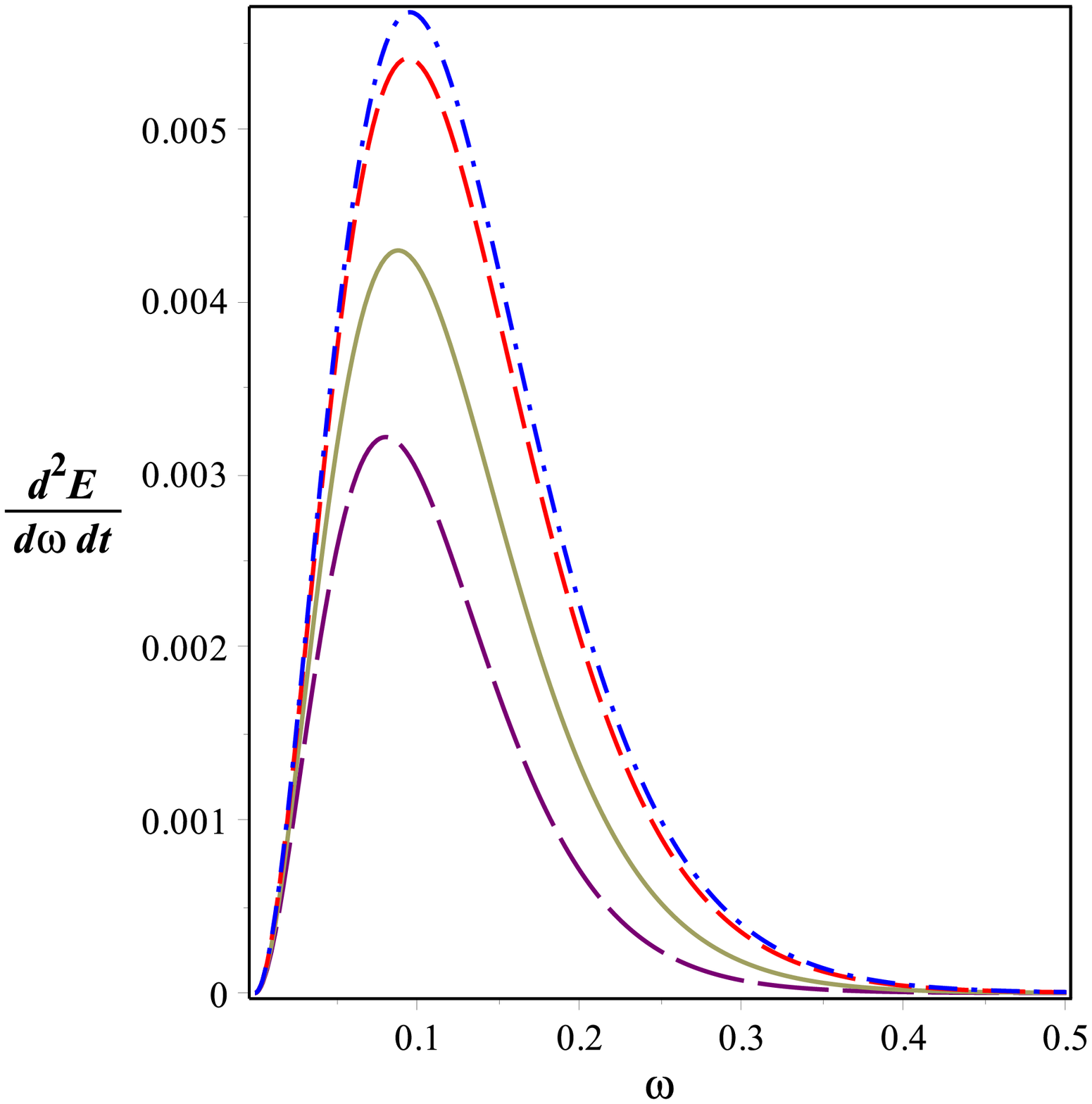}
	}
	
	\caption{\label{de}Plot showing the behavior of energy emission rate versus the frequency $\omega$ for $a=0$, $a=0.2$, $a=0.5$ and $a=0.7$ in (a), (b), (c) and (d). The value of $Q$ increases in each plot from top to down. The detail of parameters are shown in table~\ref{tab2}(see Appendix).}
 \label{pic:shadow}
\end{figure}

\clearpage

\section{shadow of black hole in the presence of plasma at $r \longrightarrow \infty$}\label{plasma}
In this section, we will analyze the shadow of Kerr-Sen dilaton-axion black hole in the presence of plasma for an observer at infinity.
The travel of radiation for an isotropic and dispersive environment in General Relativity, was studied by Bicak and Hadrava~\cite{Bick:1975} in 1975.
Furthermore, in Refs.~\cite{Er:2013efa,Morozova:2012,Tsupko:2012}, the shadow of different black hole in the presence of plasma, has been investigated. So, in this part, we analyze the influence of the plasma parameter, on the image of black hole's shadow in Kerr-Sen dilaton-axion space-time. 
The refraction index of plasma $n$, which is connected to the photon four-momentum, is~\cite{synge:11}
\begin{equation}\label{32}
n^{2}=n^{2}(x^{i},\omega)= 1 +\frac{p_{\alpha}p^{\alpha}}{(p_{\beta} u^{\beta})^{2}}
\end{equation}
where $u^{\beta}$, is the velocity for an observer, and for the vacuum environment, $n$ is equal to 1. For analytical result, we introduce the specific form of plasma frequency as ~\cite{Atamurotov:2015nra}
\begin{equation}\label{33}
n^{2}= 1-\frac{\omega^{2}_{e}}{\omega_{v}^{2}} ,
\end{equation}
where, $\omega_{v}$ and $\omega_{e}$, are photon frequency and plasma frequency, respectively. With the equation of Hamilton-Jacobi for this geometry~\cite{BisnovatyiKogan:2010ar}
\begin{equation}\label{34}
\frac{\partial S}{\partial \tau}=-\frac{1}{2}[g^{ij}p_{i}p_{j}-(n^{2}-1)(p_{0}\sqrt{-g^{00}})^{2}] . 
\end{equation}
The equations of motion of photons in the presence of plasma can be calculated as
\begin{eqnarray}\label{35}
\rho^{4}(\dfrac{dr}{d\tau})^{2}&=&-\Delta_{r}(K+\varepsilon r(r+r_{\alpha})) \nonumber\\
&+&\big((a^{2}+r(r+r_{\alpha}))E-aL \big)^{2}+(r(r+r_{\alpha})+a^{2})^{2}(n^{2}-1)E^{2}=R(r),
\end{eqnarray}
\begin{align}\label{36}
\rho^{4}(\dfrac{d\theta}{d\tau})^{2}=\Delta_{\theta}(K-\varepsilon a^{2}cos^{2}\theta)-\dfrac{1}{sin^{2}\theta}\big(aE sin^{2}\theta -L \big)^{2}-(n^{2}-1)a^{2}E^{2}\sin ^{2}\theta=\Theta(\theta),
\end{align}
\begin{align}\label{37}
\rho^{2}(\dfrac{d\varphi}{d\tau})=\dfrac{a}{\Delta_{r}}\big((a^{2}+r(r+r_{\alpha})) E - aL \big)-\dfrac{1}{sin^{2}\theta}(a E sin^{2}\theta - L),
\end{align}
\begin{align}\label{38}
\rho^{2}(\dfrac{dt}{d\tau})=\dfrac{(a^{2}+r(r+r_{\alpha}))}{\Delta_{r}}\big((a^{2}+r(r+r_{\alpha})) n^{2}E - aL \big) - a (a n^{2} E sin^{2}\theta - L).
\end{align}
We consider the plasma frequency as~\cite{Abdujabbarov:2015pqp}
\begin{equation}\label{39}
\omega^{2}_{e}=\frac{4\pi e^{2}N(r)}{m_{e}},
\end{equation}
where, $N(r)$, is the plasma number density. Also, in Eq.~(\ref{39}) $m$ and $e$ are mass and electron charge. We can consider the plasma number density as below
\begin{equation}\label{40}
N(r) = \frac{N_{0}}{r^{h}} ,
\end{equation}
So, we have 
\begin{equation}
\omega_{e}^{2}=\frac{4\pi e^{2}N_{0}}{m_{e}r^{h}},
\end{equation}
in which, $h\geq0$. In following, for this case, we consider $h=1$~\cite{Rogers:2015dla} and $\frac{4\pi e^{2}N_{0}}{m_{e}}=k$, so
\begin{equation}
\omega_{e}^{2}=\frac{k}{r},
\end{equation}
and
\begin{equation}
n=\sqrt{1-\frac{k}{r}}.
\end{equation}
In following, we obtain the constants of motion (i.e. $\eta$ and $\xi$), using $R(r)=0$ and $\dot{R}(r)=0$ conditions in Eq.~(\ref{35}). Then, for an observer in equatorial plane $\theta_{o}$=$\pi/2$, the celestial coordinates ~(\ref{alpha})--~(\ref{beta}) are
\begin{eqnarray}\label{42}
\alpha &=&-\frac{\xi}{n} ,\nonumber \\                     
\beta &=&\frac{\sqrt{\eta+a^{2}-n^{2}a^{2}}}{n} .
\end{eqnarray}
Now, some examples of black hole's shadow in the presence of plasma are plotted by using $\alpha$ and $\beta$, in Fig.~\ref{fig9}. In this figure, we show that, the shape and the size of shadows are dependent to plasma parameter, spin and $Q$.
One can see that, the asymmetry of black hole's shadow increases by increasing $a$ and the size of shadow decreases by increasing $Q$, which these influences are similar to the vacuum state.
\begin{figure}[h]\label{fig9}
	\centering
	\subfigure[$a=0$]{
		\includegraphics[width=0.4\textwidth]{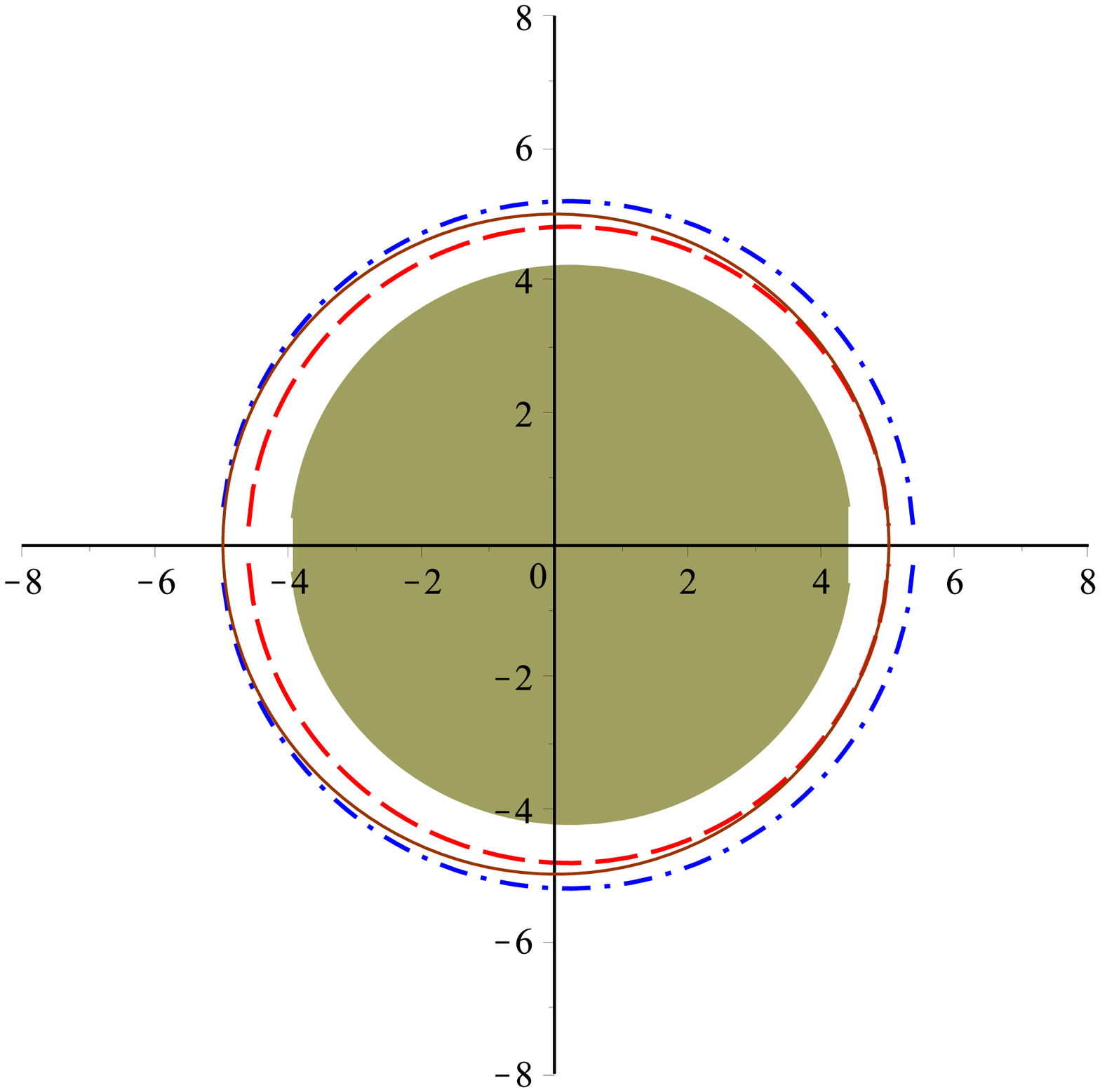}
	}
	\subfigure[$a=0.5$]{
		\includegraphics[width=0.4\textwidth]{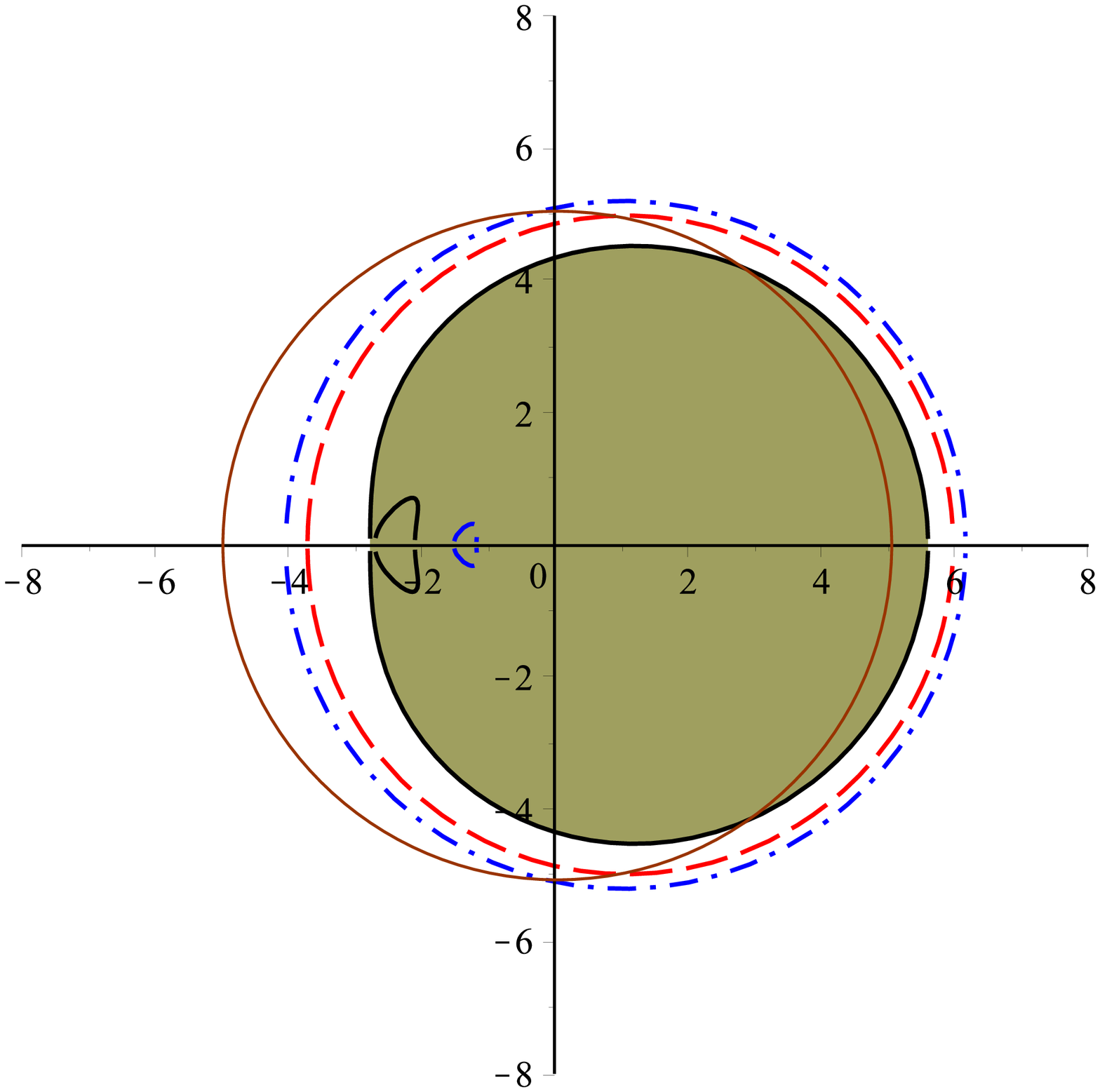}
	}
	\subfigure[$a=0.7$]{
		\includegraphics[width=0.4\textwidth]{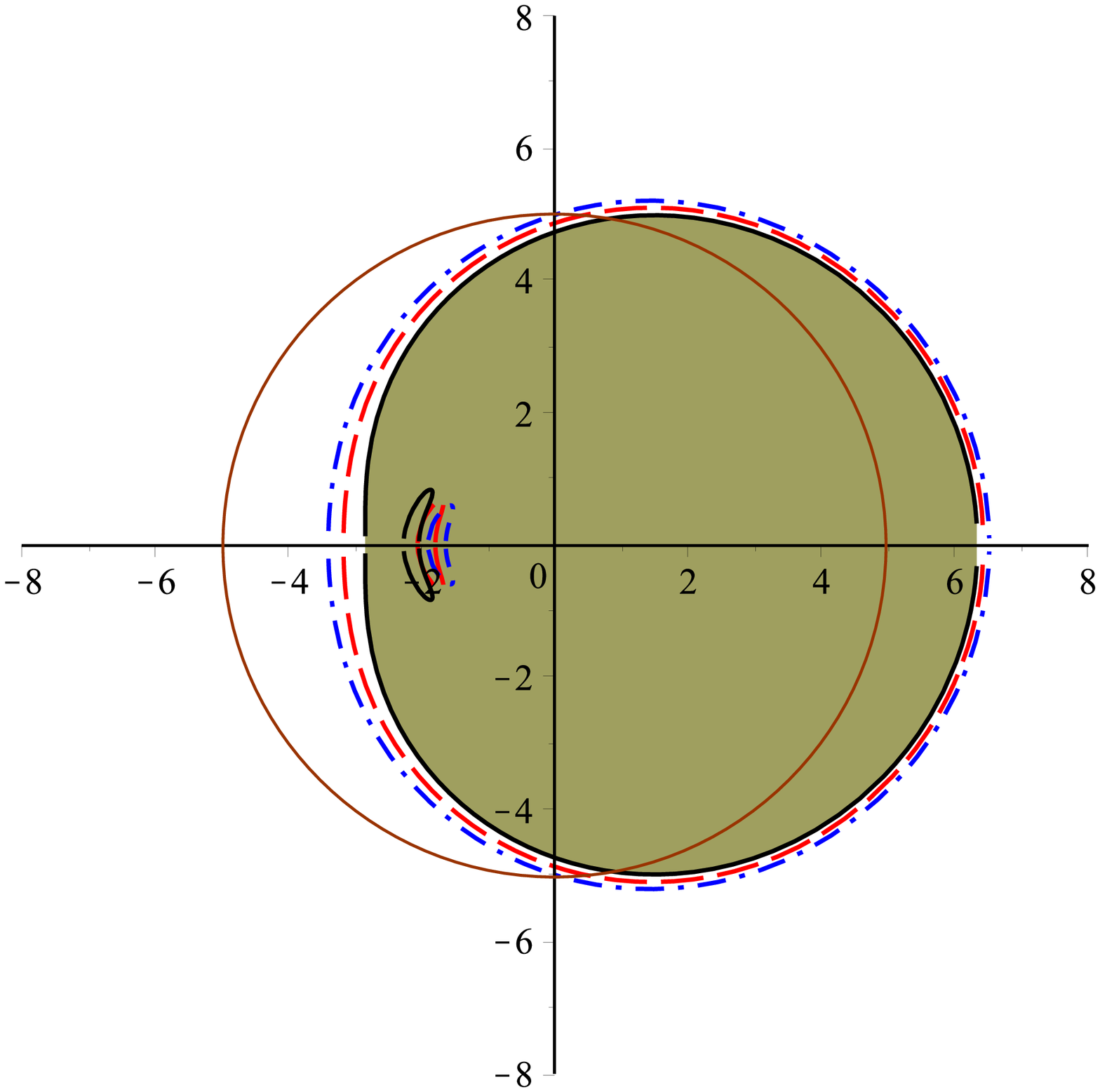}
		}
		\subfigure[$a=0.8$]{
		\includegraphics[width=0.4\textwidth]{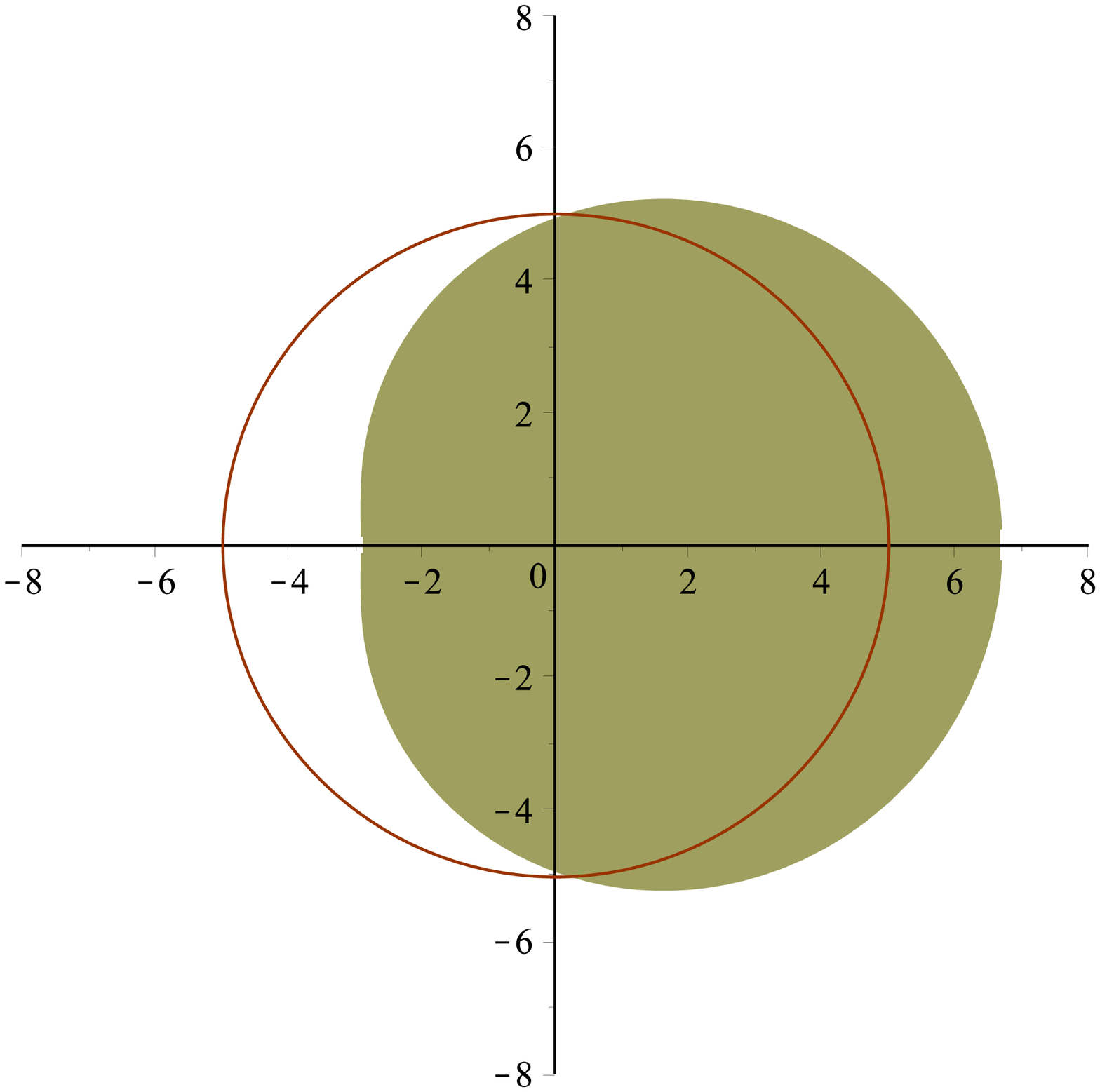}
		}
	\caption{\label{fig9}Shadow of the black hole in the presence of plasma for different values of  rotation parameter, $a=0$, $a=0.5$, $a=0.7$ and $a=0.8$ in (a), (b), (c) and (d) respectively. In (a), (b) and (c) $Q=0$, $Q=\frac{Q_{crit}}{2}$ and $Q\simeq Q_{crit}$ is shown by blue dashed-dotted line, red dashed line and green solid filled line respectively but in (d), $Q=0.1$ is plotted. The plasma parameter in these figures is $k=0.15$. The solid brown circle is reference circle. The detail of parameters are shown in table~\ref{tab3}(see Appendix).}
 \label{pic:shadow}
\end{figure}
\clearpage
In Fig.~\ref{fig11}, the effect of plasma parameter $k$, has been investigated on the size of shadow. We show that in the presence of plasma, the size of black hole's shadow decreases by increasing $k$.
Note that, the spin parameter $a$, has an important role in these changes. It means, there is no difference between image of shadow in vacuum and in the presence of plasma when $a=0$, but this difference appears by increasing $a$.
\begin{figure}[h]\label{fig11}
	\centering
	\subfigure[$a\simeq0$]{
		\includegraphics[width=0.4\textwidth]{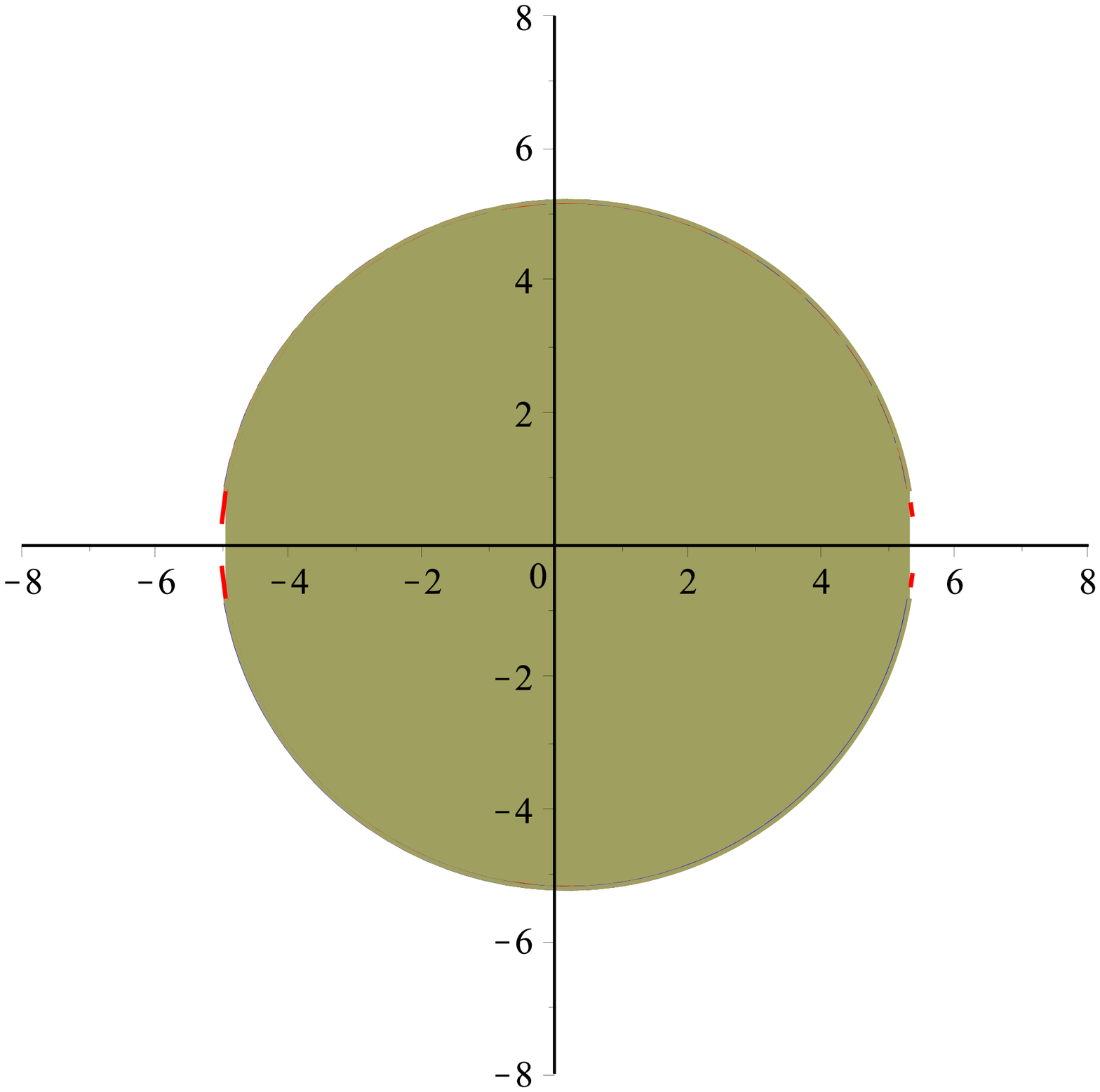}
	}
	\subfigure[$a=0.5$]{
		\includegraphics[width=0.4\textwidth]{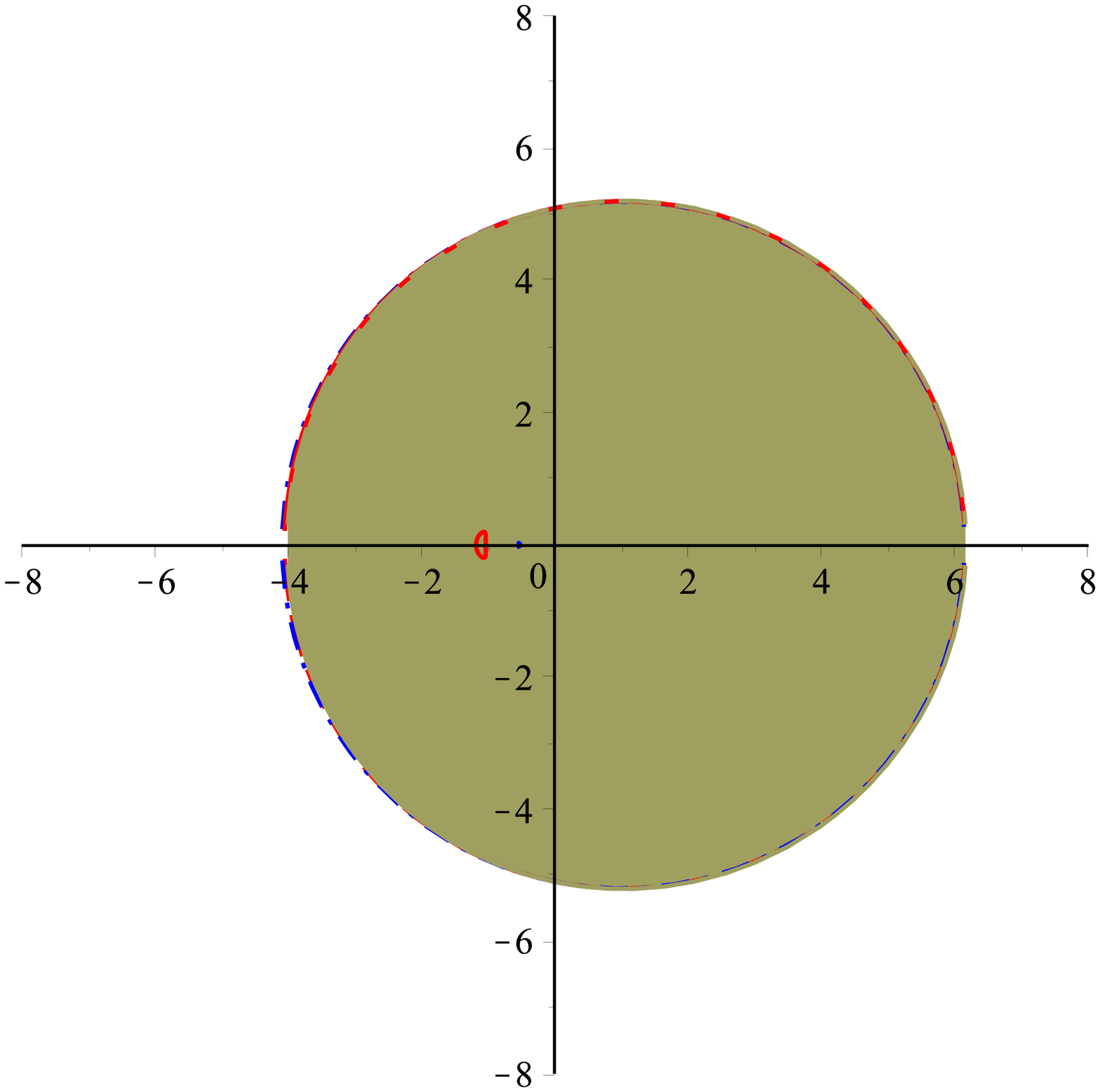}
	}
	\subfigure[$a=0.7$]{
		\includegraphics[width=0.4\textwidth]{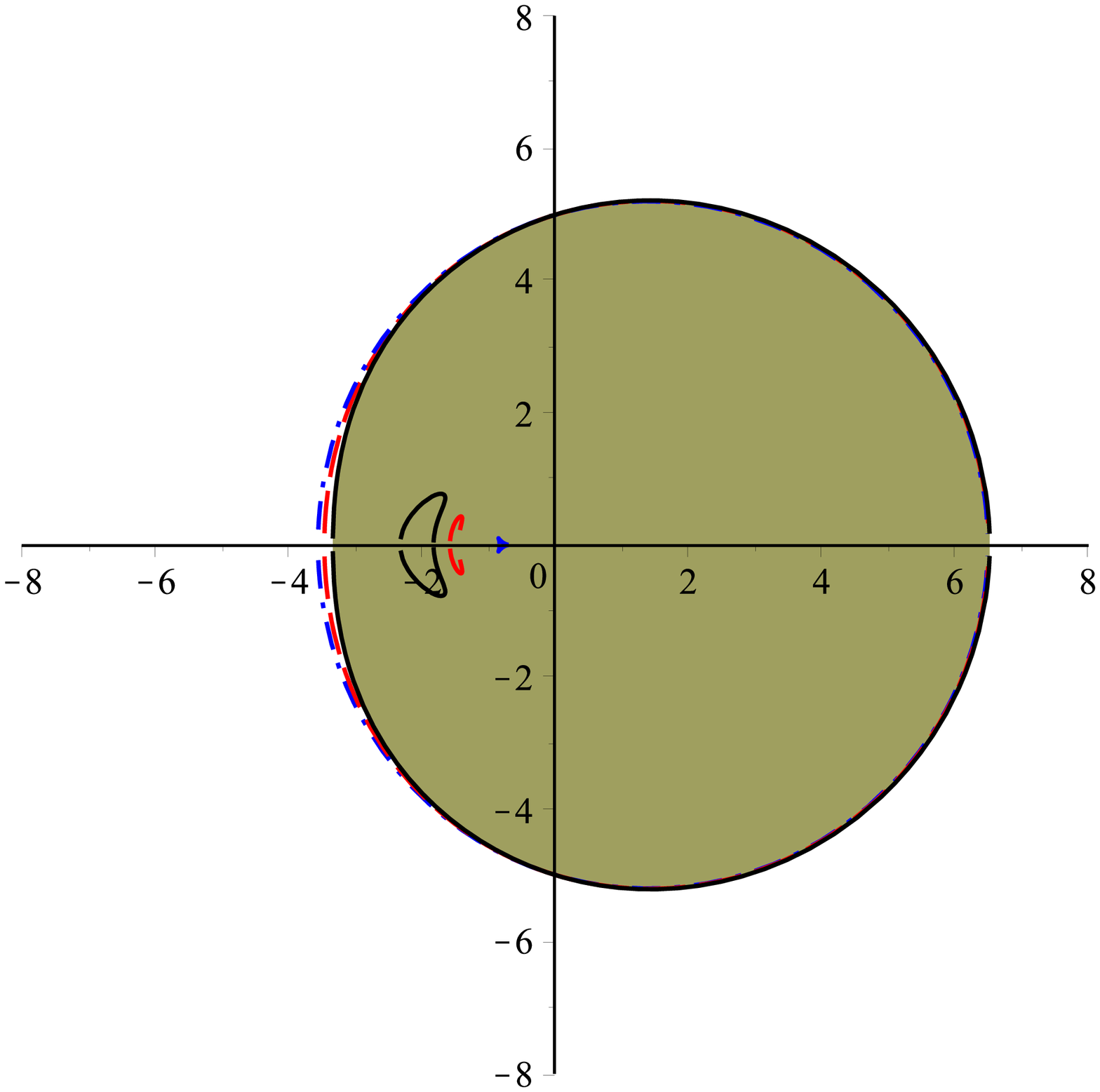}
		}
		\subfigure[$a=0.8$]{
		\includegraphics[width=0.4\textwidth]{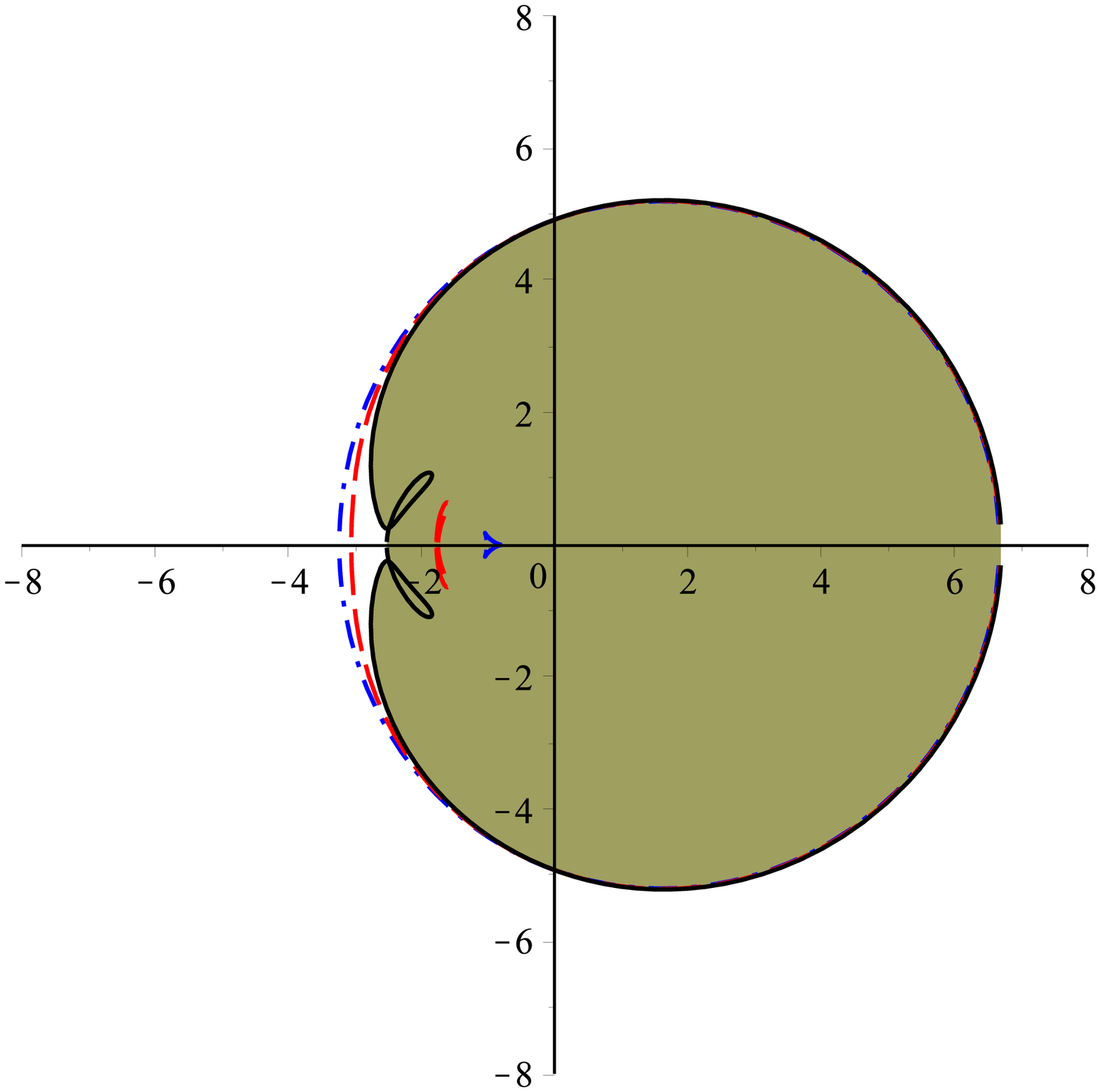}
		}
	\caption{\label{fig11}Shadow of the black hole in the presence of plasma for different values of the rotation parameter $a=0$, $a=0.5$, $a=0.7$ and $a=0.8$ in (a), (b), (c) and (d) respectively and $Q=0$. In each figure, $k=0$, $k=0.1$ and $k=0.2$ is shown by blue dashed-dotted line, red dashed line and green solid filled line respectively.}
 \label{pic:shadow}
\end{figure}
So, the radius of black hole's shadow in the presence of plasma, is always equal to or less than the vacuum state.

\clearpage
\section{CONCLUSIONS}\label{conclusion}
In this article, we studied the black hole's shadow in the vacuum state and in the presence of plasma for an observer at infinity. For analyzing the shadow, first we calculated geodesic equation for this space-time, then using geodesic equation, some example of shadow images were plotted.
The results show that, some properties of black hole's shadow, like size and symmetry, are dependent to space-time parameters. It means that, for all cases (vacuum-plasma), the size of shadow image decreases by increasing electric charge, also the symmetry of black hole's shadow deviates from circular form, by increasing $a$. Moreover, the size of black hole's shadow in vacuum is bigger than or equal to the size of shadow in the presence of plasma.
Also, the energy emission rate for different value of $a$ was plotted and we see that the energy emission rate decreases by increasing $Q$. Details of each figure, are shown in table(\ref{tab0}--\ref{tab3}) (see Appendix). 
For future research, it would be interesting to study photon region in Kerr-Sen dilaton-axion black hole.
\clearpage

\section*{Appendix}
\begin{table}[h]
\begin{center}
\begin{tabular}{|c|c|c|c|c|}

\hline
$\qquad$Constant parameters$\qquad$&Fig.\ref{infinity}&$\qquad$ $a$ $\qquad$&$\qquad$ $Q_{crit}$ $\qquad$& $\qquad$ $r_{+}- r_{-}$ $\qquad$ \\
\hline\hline

 & (a) &0 & 1.3 & 0.28,0.028\\                    
$\theta_{O}=\frac{\pi}{2}$ & (b) & 0.5 &  1 & 0.5,0.5 \\
 $M=1$& (c) & 0.7 & 0.7 & 1.03,0.47\\
 & (d) &1 & 0 & 1

\\ \hline
\end{tabular}
\caption{\label{tab0}The details of parameter in Fig.\ref{infinity}, for an observer at infinity, the electric charge and spin parameter were investigated in this figure in vacuum.}
\label{tab1}
\end{center}
\end{table}

\begin{table}[h]
\begin{center}
\begin{tabular}{|c|c|c|c|c|c|}
\hline
$\qquad$Fig.\ref{de}$\qquad$&$\qquad$$a$$\qquad$&$\qquad$ $Q$ $\qquad$&$\qquad$ $T$ $\qquad$& $\qquad$$r_{+}$$\qquad$ &$\qquad$$R_{s}$$\qquad$\\
\hline\hline

 &$\simeq 0$&0.7&$\frac{0.1380}{\pi}$&1.5&4.7\\
(a)&$\simeq 0$&1&$\frac{0.1397}{\pi}$&0.99&4\\
&$\simeq 0$&1.1&$\frac{0.1308}{\pi}$&0.77&3.8\\
&$\simeq 0$&1.2&$\frac{0.089}{\pi}$&0.54&3.6\\
\hline
 &0.2&0&$\frac{12}{100\pi}$&1.97&5.2\\
(b)&0.2&1&$\frac{11}{100\pi}$&0.95&4.1\\
&0.2&1.05&$\frac{10}{100\pi}$&0.85&3.9\\
&0.2&1.1&$\frac{7}{100\pi}$&0.73&3.8\\
 \hline
 &0.5&0&$\frac{12}{100\pi}$&1.86&5.2\\
(c)&0.5&0.4&$\frac{118}{1000\pi}$&1.69&5.05\\
&0.5&0.5&$\frac{117}{1000\pi}$&1.59&5\\
&0.5&0.6&$\frac{113}{1000\pi}$&1.46&4.85\\
 \hline
 &0.7&0&$\frac{11}{100\pi}$&1.71&5.2\\
(d)&0.7&0.2&$\frac{10}{100\pi}$&1.66&5.15\\
&0.7&0.4&$\frac{98}{1000\pi}$&1.51&5\\
&0.7&0.5&$\frac{89}{1000\pi}$&1.4&4.9\\
 \hline
\end{tabular}
\caption{\label{tab2}The details of parameter in Fig.\ref{de} .}
\label{tab2}
\end{center}
\end{table}
\begin{table}[h]
\begin{center}
\begin{tabular}{|c|c|c|c|c|}

\hline
$\qquad$Constant parameters$\qquad$&Fig.\ref{fig9}&$\qquad$ $a$ $\qquad$&$\qquad$ $Q_{crit}$ $\qquad$& $\qquad$ $r_{+}- r_{-}$ $\qquad$ \\
\hline\hline

 $k=0.15$& (a) &0 & 1 & 0.99--0.008\\                    
$\theta_{O}=\frac{\pi}{2}$ & (b) & 0.5 & 0.85 & 1.03--0.24 \\
 $M=1$& (c) & 0.7 & 0.5 & 1.4--0.35\\
 & (d) &0.8 & 0.1 & 1.58--0.40

\\ \hline
\end{tabular}
\caption{\label{tab3}The details of parameter in Fig.\ref{fig9}, for an observer at infinity, the electric charge ($Q$) and spin parameter ($a$) were investigated in this figure in the presence of plasma.}
\label{tab1}
\end{center}
\end{table}

\clearpage
\bibliographystyle{amsplain}

\begin{thebibliography}{5}
\bibitem{Wald:1984}
R. M. Wald. 
 Chicago: Univ. of Chicago Press.  (1984) 491 P.

\bibitem{Gyulchev:2009dx}
  G. N. Gyulchev and S. S. Yazadjiev,
  Phys.\ Rev.\ D {\bf 81} (2010) 023005
  doi:10.1103/PhysRevD.81.023005
  [arXiv:0909.3014 [gr-qc]].
  
\bibitem{Gyulchev:2006zg} 
  G.~N.~Gyulchev and S.~S.~Yazadjiev,
  Phys.\ Rev.\ D {\bf 75}, 023006 (2007)
  [gr-qc/0611110].


\bibitem{Antonucci:2011zza} 
  F. Antonucci {\it et al.},
  Class.\ Quant.\ Grav.\  {\bf 28}, 094001 (2011).
  
\bibitem{Doeleman:2008qh}
  S.~Doeleman {\it et al.},
  Nature {\bf 455} (2008) 78
  [arXiv:0809.2442 [astro-ph]].
  
\bibitem{EHT:2008qh}
project website:www.Eventhorizontelescope.org.

\bibitem{BHC:2008qh}
project website:BlackHoleCam.org.

\bibitem{Grenzebach:2014fha}
  A.~Grenzebach, V.~Perlick and C.~Lämmerzahl,
  Phys.\ Rev.\ D {\bf 89} (2014) no.12,  124004
  [arXiv:1403.5234 [gr-qc]].
  
\bibitem{synge}
J. L. Synge. 1966. Mon.Not.Roy.Astron.Soc.,131,463.

\bibitem{Chandrasekhar:1985kt} 
  S.~Chandrasekhar,
  OXFORD, UK: CLARENDON (1985) 646 P.
  
  \bibitem{Bardeen}
J. M. Bardeen, in Black Holes (Les Astres Occlus), edited by
C. DeWitt and B. S. DeWitt (Gordon and Breach, New York,
1973) p. 215

\bibitem{Vincent:2016sjq} 
  F.~H.~Vincent, E.~Gourgoulhon, C.~Herdeiro and E.~Radu,
  arXiv:1606.04246 [gr-qc].
  
\bibitem{Cunha:2015yba} 
  P.~V.~P.~Cunha, C.~A.~R.~Herdeiro, E.~Radu and H.~F.~Runarsson,
  Phys.\ Rev.\ Lett.\  {\bf 115}, no. 21, 211102 (2015)
  doi:10.1103/PhysRevLett.115.211102
  [arXiv:1509.00021 [gr-qc]].

\bibitem{kerr-newman}
A. de Vries, Class. Quantum Grav. 17, 123 (2000).

\bibitem{Amarilla:2011fx}
  L.~Amarilla and E.~F.~Eiroa,
  Phys.\ Rev.\ D {\bf 85} (2012) 064019
  [arXiv:1112.6349 [gr-qc]].
  
\bibitem{Amarilla:2010zq}
  L.~Amarilla, E.~F.~Eiroa and G.~Giribet,
  Phys.\ Rev.\ D {\bf 81} (2010) 124045
  [arXiv:1005.0607 [gr-qc]].
  
\bibitem{Yumoto:2012kz}
  A.~Yumoto, D.~Nitta, T.~Chiba and N.~Sugiyama,
  Phys.\ Rev.\ D {\bf 86} (2012) 103001
  [arXiv:1208.0635 [gr-qc]].
  
\bibitem{Dastan:2016vhb} 
  S.~Dastan, R.~Saffari and S.~Soroushfar,
  arXiv:1606.06994 [gr-qc].
  
  
\bibitem{Sen:1992ua} 
  A.~Sen,
  Phys.\ Rev.\ Lett.\  {\bf 69}, 1006 (1992)
  doi:10.1103/PhysRevLett.69.1006
  [hep-th/9204046].
  
\bibitem{Larranaga:2010mb} 
  A.~Larranaga,
  Pramana {\bf 76}, 553 (2011)
  doi:10.1007/s12043-011-0065-8
  [arXiv:1003.2973 [gr-qc]].
  
\bibitem{Pradhan:2015yea} 
  P.~Pradhan,
  Eur.\ Phys.\ J.\ C {\bf 76}, no. 3, 131 (2016)
  doi:10.1140/epjc/s10052-016-3976-1
  [arXiv:1503.04514 [gr-qc]].
  
\bibitem{Jiang:2006mu} 
  Q.~Q.~Jiang, S.~Z.~Yang and D.~Y.~Chen,
  Chin.\ Phys.\  {\bf 15}, 1709 (2006).
  doi:10.1088/1009-1963/15/8/013
  
\bibitem{Li:2007af} 
  G.~Q.~Li,
  Theor.\ Math.\ Phys.\  {\bf 153}, 1652 (2007).
  doi:10.1007/s11232-007-0137-6
  
\bibitem{Yang:2007ny} 
  S.~Z.~Yang and D.~Y.~Chen,
  Chin.\ Phys.\ Lett.\  {\bf 24}, 39 (2007).
  doi:10.1088/0256-307X/24/1/011
  
\bibitem{Chen:2009zzk} 
  D.~Y.~Chen and X.~T.~Zu,
  Mod.\ Phys.\ Lett.\ A {\bf 24}, 1159 (2009).
  doi:10.1142/S0217732309027133
  
\bibitem{Hioki:2008zw} 
  K.~Hioki and U.~Miyamoto,
  Phys.\ Rev.\ D {\bf 78}, 044007 (2008)
  doi:10.1103/PhysRevD.78.044007
  [arXiv:0805.3146 [gr-qc]].
  
\bibitem{Soroushfar:2016yea} 
  S.~Soroushfar, R.~Saffari and E.~Sahami,
  Phys.\ Rev.\ D {\bf 94}, no. 2, 024010 (2016)
  doi:10.1103/PhysRevD.94.024010
  [arXiv:1601.03143 [gr-qc]].
  
\bibitem{Yazadjiev:1999ce} 
  S.~Yazadjiev,
  Gen.\ Rel.\ Grav.\  {\bf 32}, 2345 (2000)
  [gr-qc/9907092].
  
  \bibitem{K}
B. Carter, Phys. Rev. {\bf 174}, 1559, (1968).

\bibitem{Wei:2013kza}
  S.~W.~Wei and Y.~X.~Liu,
  JCAP {\bf 1311} (2013) 063
  doi:10.1088/1475-7516/2013/11/063
  [arXiv:1311.4251 [gr-qc]].
  
\bibitem{Bardeen:1972fi}
  J.~M.~Bardeen, W.~H.~Press and S.~A.~Teukolsky,
  Astrophys.\ J.\  {\bf 178} (1972) 347.
  
\bibitem{Vazquez:2003zm}
  S.~E.~Vazquez and E.~P.~Esteban,
  Nuovo Cim.\ B {\bf 119} (2004) 489
  [gr-qc/0308023].
  
  
\bibitem{Hioki:2009na} 
  K.~Hioki and K.~i.~Maeda,
  Phys.\ Rev.\ D {\bf 80}, 024042 (2009)
  doi:10.1103/PhysRevD.80.024042
  [arXiv:0904.3575 [astro-ph.HE]].
  
\bibitem{Abdujabbarov:2015xqa} 
  A.~A.~Abdujabbarov, L.~Rezzolla and B.~J.~Ahmedov,
  Mon.\ Not.\ Roy.\ Astron.\ Soc.\  {\bf 454}, no. 3, 2423 (2015)
  doi:10.1093/mnras/stv2079
  [arXiv:1503.09054 [gr-qc]].
  
\bibitem{Mashhoon:1973zz} 
  B.~Mashhoon,
  Phys.\ Rev.\ D {\bf 7}, 2807 (1973).
  doi:10.1103/PhysRevD.7.2807
  
  
\bibitem{Abdujabbarov:2016hnw} 
  A.~Abdujabbarov, M.~Amir, B.~Ahmedov and S.~G.~Ghosh,
  Phys.\ Rev.\ D {\bf 93}, no. 10, 104004 (2016)
  doi:10.1103/PhysRevD.93.104004
  [arXiv:1604.03809 [gr-qc]].
  
\bibitem{Bick:1975}
  J. Bicak, P. Hadrava, \textit{Astronomy and Astrophysics} 44, 389 (1975)
  
  \bibitem{Tsupko:2012}
  O.Y. Tsupko, G.S. Bisnovatyi-Kogan, \textit{Gravitation
and Cosmology} 18, 117 (2012).


  \bibitem{Morozova:2012}
V.S. Morozova, B.J. Ahmedov, A.A. Tursunov, \textit{Astrophys Space Sci} 346, 513 (2013).

\bibitem{Er:2013efa}
  X.~Er and S.~Mao,
  Mon.\ Not.\ Roy.\ Astron.\ Soc.\  {\bf 437} (2014) no.3,  2180
  [arXiv:1310.5825 [astro-ph.CO]].
  
  
\bibitem{synge:11}
  J. L. Synge, Relativity: \textit{The General Theory}. (NorthHolland, Amsterdam, 1960)
  
\bibitem{Atamurotov:2015nra}
  F.~Atamurotov, B.~Ahmedov and A.~Abdujabbarov
  Phys.\ Rev.\ D {\bf 92} (2015) 084005
  [arXiv:1507.08131 [gr-qc]].
  
\bibitem{BisnovatyiKogan:2010ar}
  G.~S.~Bisnovatyi-Kogan and O.~Y.~Tsupko,
  Mon.\ Not.\ Roy.\ Astron.\ Soc.\  {\bf 404} (2010) 1790
  [arXiv:1006.2321 [astro-ph.CO]].
  
\bibitem{Abdujabbarov:2015pqp} 
  A.~Abdujabbarov, B.~Toshmatov, Z.~Stuchlík and B.~Ahmedov,
  arXiv:1512.05206 [gr-qc]
  
\bibitem{Rogers:2015dla}
  A.~Rogers,
  Mon.\ Not.\ Roy.\ Astron.\ Soc.\  {\bf 451} (2015) no.1,  17
  [arXiv:1505.06790 [gr-qc]].
 \end{thebibliography}

\end{document}